\documentclass[runningheads]{llncs}
\usepackage{graphicx}
\usepackage{paralist}
\usepackage{amsfonts}
\usepackage[ruled,noline,noend,linesnumbered]{algorithm2e}
\usepackage{xcolor}
\usepackage[misc,geometry]{ifsym} 

\newcommand{\algname}{\texttt{SAKEIMA}}
\newcommand{\D}{\mathcal{D}}
\newcommand{\fkm}{FK}
\newcommand{\E}{\mathbb{E}}

\begin{document}
\title{Fast Approximation of Frequent $k$-mers\\ and Applications to Metagenomics\thanks{This work is supported, in part, by the University of Padova grants \emph{SID2017} and \emph{STARS: Algorithms for Inferential Data Mining}.}
}

%
%
\author{Leonardo Pellegrina
\and
Cinzia Pizzi\and
Fabio Vandin\Letter}

\authorrunning{L. Pellegrina et al.}
%
\institute{Department of Information Engineering, University of Padova,
Padova (Italy)
\email{\{pellegri,cinzia.pizzi\}@dei.unipd.it};
\email{fabio.vandin@unipd.it}
}
\maketitle              
\begin{abstract}
Estimating the abundances of all $k$-mers in a set of biological sequences is a fundamental and challenging problem with many applications in biological analysis. While several methods have been designed for the exact or approximate solution of this problem, they all require to process the entire dataset, that can be extremely expensive for high-throughput sequencing datasets. While in some applications it is crucial to estimate all $k$-mers and their abundances, in other situations reporting only \emph{frequent} $k$-mers, that appear with relatively high frequency in a dataset, may suffice. This is the case, for example, in the computation of $k$-mers' abundance-based distances among datasets of reads, commonly used in metagenomic analyses.

In this work, we develop, analyze, and test, a sampling-based approach, called \algname, to approximate the frequent $k$-mers and their frequencies in a high-throughput sequencing dataset while providing rigorous guarantees on the quality of the approximation. \algname\ employs an advanced sampling scheme and we show how the characterization of the VC dimension, a core concept from statistical learning theory, of a properly defined set of functions leads to practical bounds on the sample size required for a rigorous approximation. Our experimental evaluation shows that \algname\ allows to rigorously approximate frequent $k$-mers by processing only a fraction of a dataset and that the frequencies estimated by \algname\ lead to accurate estimates of $k$-mer based distances between high-throughput sequencing datasets. Overall, \algname\ is an efficient and rigorous tool to estimate $k$-mers abundances providing significant speed-ups in the analysis of large sequencing datasets.

\end{abstract}

 \section{Introduction}
 \label{sec:intro}

The analysis of substrings of length $k$, called \emph{$k$-mers}, is ubiquitous in biological sequence analysis and is among the first steps of processing pipelines for a wide spectrum of applications, including: de novo assembly \cite{pevzner2001eulerian,zerbino2008velvet}, error correction \cite{kelley2010quake,salmela2016accurate}, repeat detection \cite{li2003estimating}, genome comparison \cite{sims2009alignment}, digital normalization \cite{brown2012reference}, RNA-seq quantification \cite{patro2014sailfish,zhang2014rna}, metagenomic reads classification \cite{wood2014kraken} and binning \cite{girotto2016metaprob}, fast search-by-sequence over large high-throughput sequencing repositories \cite{solomon2016fast}. A fundamental task in $k$-mer analysis is to compute the frequency of all $k$-mers, with the goal to distinguish frequent $k$-mers from infrequent $k$-mers~\cite{marccais2011fast,melsted2011efficient}. For example, this task is relevant in the analysis of high-throughput sequencing data, since infrequent $k$-mers are often assumed to result from sequencing errors. For several applications, the computation of $k$-mers frequencies is among the most computationally demanding steps of the analysis.
 
Many algorithms have been proposed for computing the exact frequency of all $k$-mers, such as Jellyfish \cite{marccais2011fast}, DSK \cite{rizk2013dsk}, KMC 3 \cite{kokot2017kmc} and Squeakr-exact \cite{pandey2017squeakr}. These methods typically perform a linear scan of the sequence to analyze, and use a combination of parallelism and efficient data structures (such as Bloom filters and Hash tables) to maintain membership and counting information associated to all $k$-mers.
Since the computation of exact $k$-mer frequencies is computationally demanding, in particular for large sequence analysis or for high-throughput sequence datasets, recent methods have focused on providing approximate solution to the problem,
improving the time and memory requirements. KmerStream \cite{melsted2014kmerstream}, khmer \cite{zhang2014these}, Kmerlight \cite{sivadasan2016kmerlight} and ntCard \cite{mohamadi2017ntcard} proposed streaming approaches for the approximation of the $k$-mer frequencies histogram. Of these, only Kmerlight and ntCard provide analytical bounds on their accuracy guarantee. KmerGenie \cite{chikhi2013informed} performs a linear scan of the input to compute the frequencies of a (random) subset of the $k$-mers that appear in the input, and uses these frequencies to approximate the abundance histogram. The recently proposed Squeakr \cite{pandey2017squeakr} relies on a probabilistic data structure to approximate the counts of individual $k$-mers. Turtle \cite{roy2014turtle} focuses on finding $k$-mers that appear at least twice in the dataset, but still processes all the $k$-mer occurrences in the input dataset, as all the other aforementioned methods do.

All the methods cited above try to estimate the frequency of \emph{all} $k$-mers or of all $k$-mers that appear at least few times (e.g., twice) in the dataset. While this is crucial in some applications (e.g., in genome assembly $k$-mers that occur exactly once often represents sequencing errors and it is therefore important to estimate the count of all observed $k$-mers), in other applications this is less justified. For example, in the comparison of high-throughput sequencing metagenomic datasets, \emph{abundance-based distances or dissimilarities} (e.g., the Bray-Curtis dissimilarity) between $k$-mer counts of two datasets are often used~\cite{benoit2016multiple,danovaro2017submarine,dickson2017carryover} to assess the distance between the corresponding datasets.  In contrast to \emph{presence-based distances}~\cite{ondov2016mash} (e.g., Jaccard distance), abundance-based distances take into account the frequency of each $k$-mer, with frequent $k$-mers contributing more to the distance than $k$-mers that appear with low frequency, but still more than a handful of times, in the dataset. 
Thus, two natural questions are \begin{inparaenum}[(i)] \item whether the results obtained considering all $k$-mers can be estimated by considering the abundances of frequent $k$-mers  only, and \item if the abundances of frequent $k$-mers can be computed more efficiently than the counts of all $k$-mers. Recently, preliminary work~\cite{hrytsenko2018efficient} has shown that, for the cosine distance and $k=12$, the answer to the first question is positive, and in Section~\ref{sec:experiments} we show that this indeed the case for larger values of $k$ and other abundance-based distances as well as presence-based distances (e.g., the Jaccard distance). To the best of our knowledge, the second question is hitherto unexplored\end{inparaenum}.  In addition, considering only frequent $k$-mers allows to focus on the most reliable information in a metagenomic dataset, since a high stochastic variability in low frequency $k$-mers is to be expected due to the sampling process inherent in sequencing.

A natural approach to reduce time and memory requirements for frequency estimation problems is to process only a portion of the data, for example by \emph{sampling} some parts of a dataset. Sampling approaches are appealing because infrequent $k$-mers naturally tend to appear with lower probability in a sample, allowing to directly focus on frequent $k$-mers in subsequent steps. However, major challenges in sampling approaches are \begin{inparaenum}[(i)]
\item to  provide rigorous guarantees relating the results obtained by processing the sample and the results that would be obtained from the whole dataset, and
\item to provide effective bounds on the size of the sample required to achieve such guarantees.
\end{inparaenum}
The application of sampling to $k$-mers is even more challenging than in other scenarios since, for values of $k$ in the typical range of interest to applications (e.g., 20-60), even the most frequent $k$-mers have relatively low frequency in the data.
To the best of our knowledge, no approach based on sampling a portion of the input dataset has been proposed to approximate frequent $k$-mers and their frequencies while providing rigorous guarantees.

\begin{figure}
  \includegraphics[width=\textwidth]{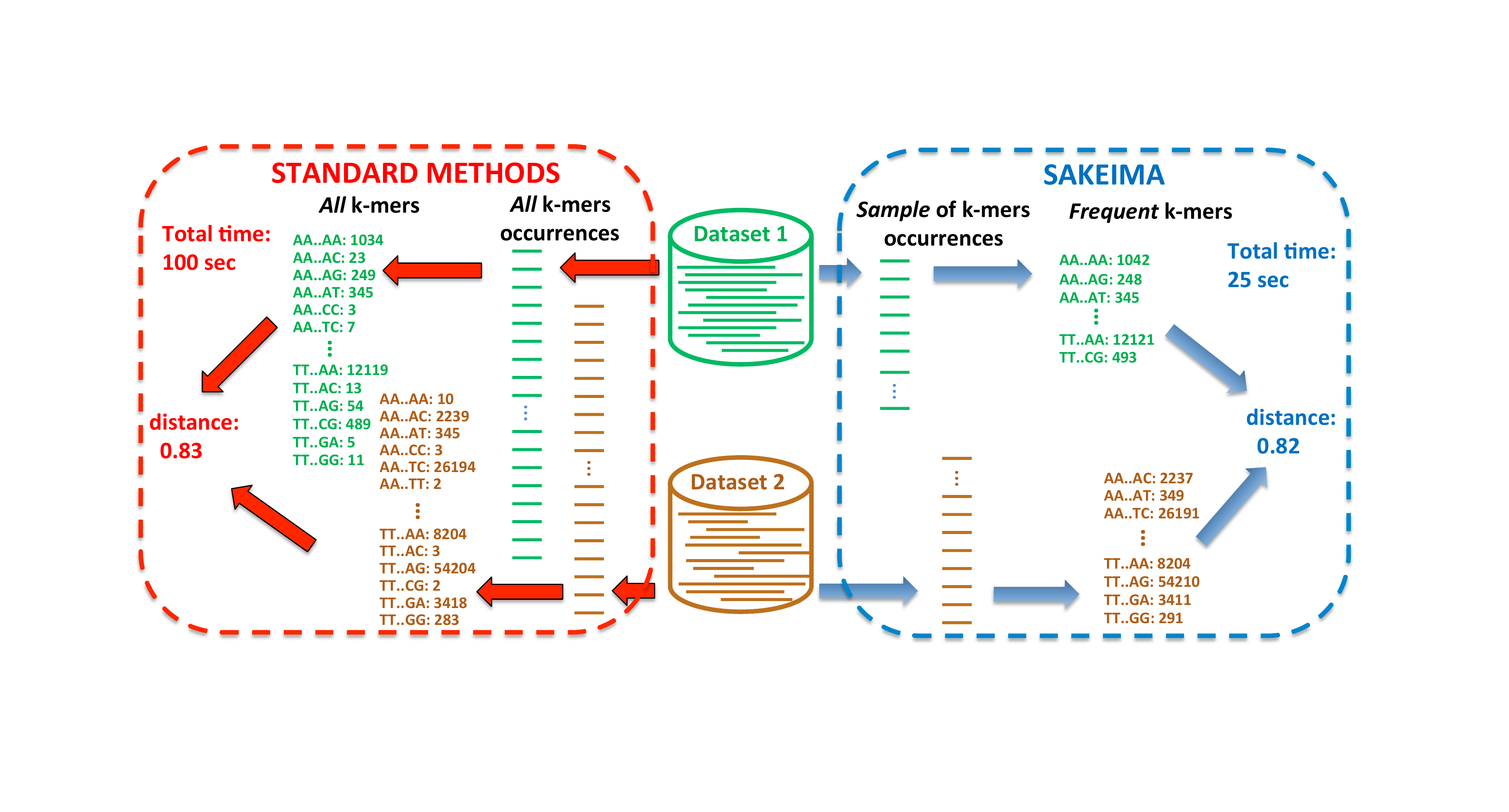}
  \caption{\algname\ computes a fast and rigorous approximation of the \emph{frequent} $k$-mers in a high-throughput sequencing dataset by sampling a fraction of all $k$-mer occurrences in a dataset, providing a significant speed-up for the computation of $k$-mer's abundance-based distances between datasets of reads (e.g., in metagenomic).}
   \label{FIG:Sakeima}
\end{figure}

\textbf{Our Contribution.} We study the problem of approximating frequent $k$-mers, i.e., $k$-mers that appear with frequency above a user-defined threshold $\theta$ in a high-throughput sequencing dataset. In these regards, our contributions are fourfold. First, we define a rigorous definition of approximation, governed by an accuracy parameter $\varepsilon$. Second, we propose a new method, \underline{S}ampling \underline{A}lgorithm for \underline{$K$}-m\underline{E}rs approx\underline{IMA}tion (\algname), to obtain an approximation to the set of frequent $k$-mers using \emph{sampling}. \algname\ is based on a sampling scheme that goes beyond na\"ive sampling of $k$-mers and allows to estimate low frequency $k$-mers considering only a fraction of all $k$-mers occurrences in the dataset. Third, we provide analytical bounds to the sample size needed to obtain rigorous guarantees on the accuracy of the estimated $k$-mer frequencies, with respect to the ones measured on the entire dataset. Our bounds are based on the notion of VC dimension, a fundamental concept from statistical learning theory. To our knowledge, ours is the first method that applies concepts from \emph{statistical learning} to provide a rigorous approximation of the $k$-mers frequencies. Fourth, we use \algname\ to extract frequent $k$-mers from metagenomic datasets from the Human Microbiome Project (HMP) and to approximate abundance-based and presence-based distances among such datasets, showing that \algname\ allows to accurately estimate such distances by analyzing only a fraction of the entire dataset, resulting in a significant speed-up.

Our approach is orthogonal to previous work: any exact or approximate algorithm can be applied to the sample extracted by \algname, that can therefore be used \emph{before} applying previously proposed methods, thus reducing their computational requirements while providing rigorous guarantees on the results w.r.t. to the entire dataset. 
While we present our methodology in the case of finding frequent $k$-mers from a set of sequences representing a high-throughput sequencing dataset of short reads, our results can be applied to datasets of long reads and to whole-genome sequences as well.

\section{Preliminaries}
\label{sec:prelims}

Let a dataset $\D$ be a bag of $n$ reads $\D =\{r_0 , \dots , r_{n-1}\}$, where
each read $r_i$, $0 \le i \le n-1$,  is a string of length $n_i$ from an alphabet $\Sigma$ of cardinality $|\Sigma|=\sigma$. For $j \in \{0,\dots,n_i-1\}$, let $r_i[j]$ be the $j$-th character of $r_i$. For a given integer $k \leq \min_i\{n_i: r_i \in \D\}$, we define a $k$-mer $A$ as a string of length $k$ from $\Sigma$, that is $A \in \Sigma^k$. We say that a $k$-mer $A$ \emph{appears} in $r_i$ at position $j \in \{0,\dots, n_i-k\}$ if $r_i[j+h]=A[h], \forall h \in \{0,\dots,k-1\}$.  For every $i, 0\le i \le n-1$, and every $j \in \{0,\dots , n_i - k\}$, we define the indicator function $\phi_{r_i,A}(j)$ that is $1$ if the $k$-mer $A$ appears in $r_i$ at position $j$, while  $\phi_{r_i,A}(j)=0$ otherwise. The total number of $k$-mers in $\D$ is $t_{\D,k} = \sum_{i=0}^{n-1} (n_i - k + 1) $. We define the \emph{support}  $o_{\D}(A)$ of a $k$-mer $A$ as the number of distinct positions in $\D$ where $A$ appears: $o_{\D}(A) = \sum_{i=0}^{n-1} \sum_{j=0}^{n_i-k} \phi_{r_i,A}(j)$. We define the \emph{frequency} $f_\D(A)$ of $A$ in $\D$ as the ratio between the number of distinct positions where $A$ appears in $\D$ and the total number of $k$-mers in $\D$: $f_\D(A) = o_{\D}(A) / t_{\D,k}$.

\subsection{Frequent $k$-mers and Approximations}

We are interested in obtaining the set $\fkm(\D,k,\theta)$ of frequent $k$-mers in a dataset $\D$ with respect to a minimum frequency threshold $\theta$, defined as follows.

\begin{definition}
Given a dataset $\D$, an integer $k>0$, and a frequency threshold $\theta \in (0,1]$, the set $\fkm(\D,k,\theta)$ of \emph{Frequent $k$-Mers in $\D$ w.r.t. $\theta$} is the collection of all $k$-mers with frequency at least $\theta$ in $\D$ and of their corresponding frequencies in $\D$:
\begin{equation}
\fkm(\D,k,\theta) = \{ (A,f_\D(A)) : f_\D(A) \geq \theta  \}.
\end{equation}
\end{definition}

$\fkm(\D,k,\theta)$ can be computed with a single scan of all the $k$-mers occurrences in $\D$ maintaining the $k$-mers supports in an appropriate data structure; however, when $\D$ is extremely large and $k$ is not small, the exact computation of $\fkm(\D,k,\theta)$ is extremely demanding in terms of time and memory, since the number of $k$-mers grows exponentially with $k$. In this case, a fast to compute \emph{approximation} of the set $\fkm(\D,k,\theta)$ may be preferable, provided it ensures rigorous guarantees on its quality. In this work, we focus on the following approximation. 

\begin{definition}
Given a dataset $\D$, an integer $k>0$, a frequency threshold $\theta \in (0,1]$, and a constant $\varepsilon \in (0,\theta)$, an \emph{$\varepsilon$-approximation} of $\fkm(\D,k,\theta)$ is a collection $C = \{ (A,f_A) : f_A \in (0, 1]  \}$ such that:
\begin{itemize}
\item for any  $(A,f_\D(A)) \in \fkm(\D,k,\theta)$  there is a pair $(A,f_A) \in C$;
\item for any $(A,f_A) \in C$ it holds that $f_\D(A) \geq \theta - \varepsilon$;
\item for any $(A,f_A) \in C$ it holds that $|f_\D(A) - f_A| \leq \varepsilon / 2$.
\end{itemize}
\end{definition}

The definition  above guarantees that every frequent $k$-mer of $\D$ is in the approximation and that no $k$-mer with frequency $<\theta - \varepsilon$ is in the approximation. The third condition guarantees that the estimated frequency $f_A$ of $A$ in the approximation is close (i.e, within $\varepsilon/2$) to the frequency $f_\D(A)$ of $A$ in $\D$. It is easy to show that obtaining a $\varepsilon$-approximation of $\fkm(\D,k,\theta)$ with absolute certainty requires to process all $k$-mers in $\D$.

\subsection{Simple Sampling-Based Algorithms and Bounds}
\label{sec:simple_bounds}

We aim to provide an approximation to $\fkm(\D,k,\theta)$ with \emph{sampling}, by processing only \emph{randomly selected portions} of $\D$. The simplest sampling scheme is the one in which a random sample is a bag $P$ of $m$ positions taken uniformly at random, with replacement, from the set $P_{\D,k} = \{(i,j):  i \in [0,n-1], j \in [0,n_i-k] \}$  (note that $|P_{\D,k}| = t_{\D,k}$) of all positions where $k$-mers occurs in the dataset $\D$, corresponding to $m$ 
occurrences of $k$-mers (with repetitions) taken uniformly at random. Given such sample $P$, an integer $k>0$, and a minimum frequency threshold $\theta \in (0,1]$ one can define the set of frequent $k$-mers (and their frequencies) in the sample $P$ as
$\fkm(P,k,\theta) = \{ (A,f_P(A)): f_P(A) \geq \theta \}$, where $f_P(A)$ is the frequency of $k$-mer $A$ in the sample.

Obtaining a $\varepsilon$-approximation from a random sample with absolute certainty is impossible, thus we focus on obtaining a $\varepsilon$-approximation with probability $1-\delta>0$, where $\delta \in (0,1)$ is a \emph{confidence} parameter, whose value is provided by the user. 
Intuitively, the set $\fkm(\D,k,\theta)$ of frequent $k$-mers  is well approximated by the set of frequent $k$-mers in a random sample $P$ when $P$ is sufficiently large. One natural question regards how many samples are needed to obtain the desired $\varepsilon$-approximation. 
By using Hoeffding's inequality~\cite{mitzenmacher2017probability} to bound the deviation of the frequency of a $k$-mer $A$ in the sample from $f_{\D}(A)$ and a union bound on the maximum number $\sigma^k$ of $k$-mers, where $\sigma = |\Sigma|$, we have the following result that provides a first such bound, and a corresponding first algorithm to obtain a $\varepsilon$-approximation to $\fkm(\D,k,\theta)$.  (Due to space constraints proofs are omitted and will be provided in the full version of this extended abstract.)

\begin{proposition}
\label{prop:unionboundsample1}
Consider a sample $P$ of size $m$ of $\D$. If
$m \geq \frac{2}{\varepsilon^2} \left( \ln \left( 2 \sigma^k\right)+\ln\left(\frac{1}{\delta} \right) \right)$ for fixed $\varepsilon \in (0,\theta), \delta \in (0,1)$,
then, with probability $\ge 1- \delta$, $\fkm(P,k,\theta-\varepsilon/2)$ is a $\varepsilon$-approximation of $\fkm(\D,k,\theta)$. 
\end{proposition}

In addition, by using known results in statistical learning theory~\cite{vapnik1971uniform,mitzenmacher2017probability} relating the VC dimension (see Section~\ref{sec:practicalbounds} for its definition) of a family of functions and a novelly derived bound on the family of functions $\{f_\D(A)\}$, we obtain  the following improved bound and algorithm. (The derivation will be provided in the full version.)

\begin{proposition}
\label{kmsamplesize1}
Let $P$ be a sample of size $m$ of $\D$. For fixed $\varepsilon \in(0,\theta), \delta \in (0,1)$, if 
$m \geq  \frac{2}{\varepsilon^2}\left(1 + \ln \left(\frac{1}{\delta}\right)\right)$
then $\fkm(P,k,\theta - \varepsilon / 2)$ is an $\varepsilon$-approximation for $\fkm(\D,k,\theta)$ with probability $\ge 1 - \delta$.
\end{proposition}

\section{Advanced and Practical Bounds and Algorithms for $k$-mer Approximations}
\label{sec:practicalbounds}

While the bound of Proposition~\ref{kmsamplesize1} significantly improves the simple bounds of Section~\ref{prop:unionboundsample1}, since the factor $\ln(2 \sigma^k)$ has been reduced to $1$, it still has an inverse quadratic dependency with respect to the accuracy parameter $\varepsilon$, that is problematic when the quantities to estimate are small. In these cases, one needs a small $\varepsilon$ to produce a meaningful approximation (since $\varepsilon < \theta$), and the inverse quadratic dependence of the sample size from $\varepsilon$ often results in a sample size larger than the entire input, defeating the purpose of sampling. The case of $k$-mers is particularly challenging, since the sum $\sum_{A \in \Sigma^k} f_\D(A)$ of all $k$-mers frequencies is exactly $1$. Therefore the higher the number of distinct $k$-mers appearing in the input, the lower their frequencies will be, with the consequence that $\theta$ (and therefore $\varepsilon$) typically needs to be set to a very low value. For example, a typical dataset from the Human Microbiome Project (HMP) has $n \approx 10^8$ reads of (average) length $\approx 100$: therefore if we are interested in $k$-mers for $k=31$, by setting $\delta=0.05$  the bound of Section~\ref{sec:simple_bounds} gives $\varepsilon \approx 10^{-5}$, that is only $k$-mers with frequency $\ge 10^{-5}$ could be reliably reported by sampling. However, in datasets we considered, no or a very small number ($\le 30$) of $k$-mers have frequency $\ge 10^{-5}$,
therefore according to the result from Section~\ref{sec:simple_bounds} we cannot obtain a meaningful approximation of $k$-mers and their frequencies.
In the remaining of this section we develop more refined sampling schemes and estimation techniques leading to a practical sampling-based algorithm.

\subsection{Sampling Bags of Positions and VC dimension Bound.}
\label{sec:sample_bags}

We propose a method to provide an efficiently computable approximation to $\fkm(\D,k,\theta)$ when the minimum frequency $\theta$ is low, 
by properly defining samples so that any $k$-mer $A$ will appear in a sample with probability higher than $f_{\D}(A)$, thus lessening the the dependence of the sample size from $1/\varepsilon^2$.
For this to be achievable, we need to relax the notion of approximation defined in Section~\ref{sec:prelims}. In particular, the guarantees, provided by our method, in such relaxed approximation are that \emph{all} $k$-mers with frequency above $\theta'$, with $\theta'$ slightly higher than $\theta$, are reported in output, and that no $k$-mer having frequency below $\theta - \varepsilon$ is reported in output.  (See Proposition~\ref{thm:kmsamplesizevellpositions} for the definition of $\theta'$.)
 Our experiments show that the fraction of $k$-mers having frequency $\in [\theta , \theta')$ which are non reported is very small. Our method works by sampling \emph{bags of positions} instead than single positions. In particular, an element of the sample is now a set of $\ell$ positions chosen independently at random from the set $P_{\D,k}$ of all positions.
 
Let $I_{\ell} = \{(i_1,j_1) , (i_2,j_2) , \dots , (i_\ell,j_\ell) \}$ be a bag of $\ell$ positions for $k$-mers in $\D$, chosen uniformly at random from the set $P_{\D,k}$.
We define the indicator functions $\hat{\phi}_{A}(I_{\ell})$ that, for a given bag $I_{\ell}$ of $\ell$ positions, is equal to $1$ if $k$-mer $A$ appears in \emph{at least} one of the $\ell$ positions in $I_{\ell}$ and is equal to $0$ otherwise. That is
$\hat{\phi}_{A}(I_{\ell})  = \min \left\{ 1 , \sum_{(i,j) \in I_{\ell}} \phi_{r_i,A}(j) \right\}.$
We define the \emph{$\ell$-positions sample $P_\ell$} as a bag of $m$ bags $\{I_{\ell,0},I_{\ell,1},\dots,I_{\ell,m-1}\}$, where each $I_{\ell,j}, 0 \le j\le m-1$ is a bag of $\ell$ positions, sampled independently, and

\begin{equation}
\hat{f}_{P_\ell}(A) = \frac{1}{m} \sum_{I_{\ell,i} \in P_\ell} \frac{ \hat{\phi}_{A}(I_{\ell,i}) }{\ell}.
\end{equation}

Intuitively, $\hat{f}_{P_\ell}(A)$ is the biased version of the unbiased estimator 
$f_{P_\ell}(A) = \frac{1}{m} \sum_{I_{\ell,i} \in P_\ell} \frac{ \sum_{(i,j) \in I_{\ell,i}} \phi_{r_i,A}(j) }{\ell}$
of $f_\D(A)$, where the bias arises from considering a value of $1$ every time \mbox{$\sum_{(i,j) \in I_{\ell,i}} \phi_{r_i,A}(j) > 1$}.

In our analysis we use the Vapnik-Chervonenkis (VC) dimension \cite{vapnik1998statistical,vapnik1971uniform}, a statistical learning concept that measures the expressivity of a family of binary functions. We define a range space $Q$ as a pair $Q = (X,R_X)$ where $X$ is a finite or infinite set and $R_X$ is a finite or infinite family of subsets of $X$. The members of $R_X$ are called \emph{ranges}. Given $D \subset X$, the \emph{projection} of $R_X$ on $D$ is defined as $proj_{R_X}(D) = \{ r \cap D : r \in R_X \}$. We say that $D$ is \emph{shattered by $R_X$} if $proj_{R_X}(D) = 2^{|D|}$. The \emph{VC dimension of $Q$}, denoted as $VC(Q)$, is the maximum cardinality of a subset of $X$ shattered by $R_X$. If there are arbitrary large shattered subsets of $X$ shattered by $R_X$, then $VC(Q)=\infty$.

A finite bound on the VC dimension of a range space $Q$ implies a bound on the number of random samples required to obtain a good approximation of its ranges, defined as follows.

\begin{definition}
\label{def:rangeepsapprox}
Let $Q=(X,R_X)$ be a range space and let $D$ be a finite subset of $X$. For $\varepsilon \in (0,1]$, a subset $B$ of $D$ is an \emph{$\varepsilon$-approximation of $D$} if for all $r \in R_X$ we have:
$\left|\frac{|D\ \cap\ r|}{|D|} - \frac{|B\ \cap\ r|}{|B|} \right| \leq \varepsilon/2.$
\end{definition}

The following result \cite{mitzenmacher2017probability} relates $\varepsilon$ and the probability that a random sample of size $m$ is an $\varepsilon$-approximation for a range space of VC dimension at most $v$.

\begin{proposition}[\cite{mitzenmacher2017probability}]
\label{prop:VCsamplesize}
There is an absolute positive constant $c$ such that if $(X,R_X)$ is a range-space of VC dimension at most $v$, $D$ is a finite subset of $X$, and $0 < \varepsilon$, $\delta < 1$, then a random subset $B \subset D$ of cardinality $m$ with
$m \geq \frac{4c}{\varepsilon^2}\left(v + \ln \left(\frac{1}{\delta}\right)\right)$
is a $\varepsilon$-approximation of $D$ with probability at least $1 - \delta$.
\end{proposition}

The universal constant $c$ has been experimentally estimated to be at most $0.5$ \cite{loffler2009shape}.

We now prove an upper bound to the VC dimension $VC(Q)$ of the range space $Q$ associated to the class of functions $\hat{\phi}_{A}$ that grows sub-linearly with respect to $\ell$. To this aim, we first define the range space associated to bags of $\ell$ positions of $k$-mers.

 \begin{definition}
\label{def:range2}
Let $\D$ be a dataset of $n$ reads and let $k$ and $\ell$ be two integers $\ge 1$. We define $Q = (X_{\D,k,\ell},R_{\D,k,\ell})$ to be the following range space:
\begin{itemize}
\item $X_{\D,k,\ell}$ is the set of all bags of $\ell$ positions of $k$-mers in $\D$, that is the set of all possible subsets, with repetitions, of size $\ell$ from  from $P_{\D,k}$;
\item $R_{\D,k,\ell} = \{P_{\D, \ell }(A) | A \in \Sigma^k \} $ is the family of sets of starting positions of $k$-mers, such that for each $k$-mer $A$, the set $P_{\D, \ell }(A)$ is the set of all bags of $\ell$ starting positions in $\D$ where $A$ appears at least once.
\end{itemize}
\end{definition}
 
We prove the following results on the VC dimension of the above range space.
 
\begin{proposition}
\label{thm:vcdiml}
Let $Q$ the range space from Definition~\ref{def:range2}. Then: $VC(Q) \leq \lfloor \log_2(\ell) \rfloor + 1$.
\end{proposition}

Using the result above, we prove the following.

\begin{proposition}
\label{thm:kmsamplesizevellpositions}
Let $\ell \ge 1$ be an integer and $P_\ell$ be a bag of $m$ bags of $\ell $ positions of $\D$ with
\begin{equation}
m \geq \frac{2}{(\ell \varepsilon)^2}\left(\lfloor \log_2\min( 2 \ell , \sigma^k ) \rfloor + \ln \left(\frac{1}{\delta}\right)\right).
\end{equation}
Then, with probability at least $1 - \delta$:
\begin{itemize}
\item for any $k$-mer $A \in \fkm(\D,k,\theta)$ such that  $f_\D(A) \ge \theta'= 1 - (1-\ell \theta)^{1/\ell}$ it holds $\hat{f}_{P_\ell}(A)  \geq \theta - \varepsilon / 2$;
\item for any $k$-mer $A$ with $\hat{f}_{P_\ell}(A) \geq \theta - \varepsilon / 2$ it holds $f_\D(A) \ge \theta - \varepsilon $;
\item for any $k$-mer $A \in \fkm(\D,k,\theta)$ it holds  $f_\D(A) \ge \hat{f}_{P_\ell}(A) - \varepsilon / 2$;
\item for any $k$-mer $A$ with $\hat{f}_{P_\ell}(A) - \varepsilon/2 \ge 0$, it holds $f_\D(A) \ge 1- ( 1 - \ell (\hat{f}_{P_\ell}(A) - \varepsilon/2))^{1 / \ell}$;
\item for any $k$-mer $A$ with $\ell (\hat{f}_{P_\ell}(A) + \varepsilon/2) \le 1$ it holds $f_\D(A) \leq 1- ( 1 - \ell (\hat{f}_{P_\ell}(A) + \varepsilon/2))^{1 / \ell}$.
\end{itemize}
\end{proposition}

Note that from Proposition~\ref{thm:kmsamplesizevellpositions} the set $\{(A,f_{P_\ell}(A)): \hat{f}_{P_\ell}(A)\geq \theta - \varepsilon / 2 \}$ is \emph{almost} a $\varepsilon$-approximation to $\fkm(\D,k,\theta)$: in particular, there may be $k$-mers $A$ for which $\E[\hat{f}_{P_\ell}(A)]=  (1 - \left( 1 - f_\D(A) \right)^\ell) / \ell < \theta$ while $f_\D(A) = \E[f_{P_\ell}(A)] \ge \theta$ and such that for the given sample $P_\ell$ we have $\hat{f}_{P_\ell}(A) \approx \E[\hat{f}_{P_\ell}(A)] - \varepsilon/2$. 
While this can happen, we can limit the probability of this happening by appropriately choosing $\ell$, and still enjoy the reduction in sample size of the order of $\frac{\log_2 \ell}{\ell^2}$ w.r.t. Proposition \ref{kmsamplesize1} obtained by considering bags of bags of $\ell$ positions.
In particular, this result allows the user to set $\theta$, $\varepsilon$, $\delta$, and $\ell$ to effectively find, with probability at least $1 - \delta$, \emph{all} frequent $k$-mers $A$ for which $f_\D(A) \ge \theta'$ and do not report any $k$-mer with frequency below $\theta - \varepsilon$, while still being able to report in output almost all $k$-mers with frequency $\in [\theta , \theta')$. Our experimental analysis (Section~\ref{sec:experiments}) shows that in practice choosing $\ell$ close from below to $1/\theta$ is very effective to obtain such result.
Then, the third, fourth, and fifth guarantees from Proposition~\ref{thm:kmsamplesizevellpositions} state that we can use the biased estimates $\hat{f}_{P_\ell}(A)$ to derive \emph{guaranteed upper and lower bounds} to $f_\D(A)$ that will be much tighter than the one obtained using the bounds of Section \ref{sec:simple_bounds}. We will show how to obtain further improved upper and lower bounds to $f_\D(A)$ in Section~\ref{sec:lower_upper_bounds}. Such lower bounds $\ell b_A$ can be used, for example, to prove that the set $\{(A,f_{P_{\ell}}(A)): \ell b_A \ge \theta - \varepsilon \}$ enjoys the same last four guarantees 
from Proposition~\ref{thm:kmsamplesizevellpositions} while the first one holds for a $\theta' < 1 - (1-\ell \theta)^{1/\ell}$; therefore, when false negatives are problematic, the set $\{(A,f_{P_{\ell}}(A)): \ell b_A \ge \theta - \varepsilon \}$ can be used to obtain a different approximation of $\fkm(\D,k,\theta)$ with fewer false negatives.

\subsection{\algname: An Efficient Algorithm to Approximate Frequent $k$-mers}

{\sloppy We now present our \underline{S}ampling \underline{A}lgorithm for \underline{K}-m\underline{E}rs approx\underline{IMA}tion (\algname), that builds on Proposition~\ref{thm:kmsamplesizevellpositions} and efficiently samples  a bag $P_\ell$ of bags of $\ell$-positions from $\D$ to obtain an approximation of the set $\fkm(\D,k,\theta)$ with probability $1 - \delta$, where $\delta$ is a parameter provided by the user.
}
\begin{figure}
\begin{algorithm}[H]
\caption{{\algname}}
\label{algo:ellpositions}
\KwIn{dataset $\D$, total number of $k$-mers $t_{\D,k}$ in $\D$,\\frequency threshold $\theta$, accuracy parameter $ \varepsilon \in (0,\theta)$,\\confidence parameter $\delta \in (0,1)$, integer $\ell \ge 1$.}
\KwOut{approximation $\{(A,f_A)\}$ of $\fkm(\D,k,\theta)$ with probability $\ge 1 - \delta$}
$m \leftarrow \left\lceil \frac{2}{(\ell \varepsilon)^2}\left(\lfloor \log_2\min( 2 \ell , \sigma^k ) \rfloor + \ln \left(\frac{2}{\delta}\right)\right)\right\rceil$;  $\lambda \gets \frac{m\ell}{t_{\D,k}}$\;
$T \gets $ empty hash table\;
\ForAll{reads $r_i \in \D$}{
	\ForAll{$j \in [0, n_i-k]$}{
		$A \gets $ $k$-mer in position $j$ of read $r_i$\;
		$a \gets  Poisson(\lambda)$\;
		\textbf{if} $a > 0$ \textbf{then }$T[A] \gets T[A] + a$\;
	}
}
$\mathcal{O} \leftarrow \emptyset $; $t \leftarrow \sum_{A \in T}T[A] $\;
$P_{\ell} \leftarrow$ random partition of $t$ occurrences in $T$ into $m$ bags\;
\ForAll{$k$-mers $A \in T$}{
	$f_A  \leftarrow T[A] / t$\;
	$\mathcal{P}_A \leftarrow$ bags of $P_{\ell}$ where $A$ appears at least once\;
	$\hat{f}_A  \leftarrow  |\mathcal{P}_A| / m$\;
	\lIf{$\hat{f}_A \ge \theta - \varepsilon/ 2$}{$\mathcal{O}  \leftarrow \mathcal{O} \cup (A, f_A)$}
}
\textbf{return} $\mathcal{O}$;
\end{algorithm}
\end{figure}

\algname\ is described in Algorithm~\ref{algo:ellpositions}.
\algname\ performs a pass on the stream of $k$-mers appearing in $\D$, and for each position in the stream it samples the number $a$ of times that the position appears in the sample $P_\ell$ independently at random from the Poisson distribution $Poisson(\lambda)$ of parameter $\lambda = m \ell/t_{\D,k}$. \algname\ stores such values in a counting structure $T$ (lines 3-7) that keeps, for each $k$-mer $A$, the total number of occurrences of $A$ in the sample $P_\ell$. 
(Note that $t_{\D,k}$, that can be computed with a very quick linear scan of the dataset, where $n_i$ is computed for every $r_i \in \D$ without extracting and processing (e.g., inserting or updating information for) $k$-mers; in alternative a lower bound to $t_{\D,k}$ can be used, simply resulting in a number of samples higher than needed). Then, such occurrences are partitioned into the $m$ bags $I_{\ell,0},\dots,I_{\ell,m-1}$  (line 9); this can be efficiently implemented by assigning each occurrence to a random bag while keeping the difference between the final size of the bags $\le 1$. For each $k$-mer $A$ appearing at least once in the sample (line 10), the unbiased estimate $f_A$ is computed as the number $T[A]$ of occurrences of $A$ in the sample $P_{\ell}$ (line 11) divided by the total number of positions in the sample, while the  biased estimate $\hat{f}_A$ is computed
as the number $|\mathcal{P}_A|$ of distinct bags  of $P_\ell$ where $A$ appears at least once divided by the number $m$ of bags (lines 12-13).
Then \algname\ flags $A$ as frequent if $\hat{f}_A \geq \theta - \varepsilon/2$ (line 14) and, in this case, the couple $(A,f_A)$ is added to the output set $\mathcal{O}$ (line 14), since $f_A$ is the best (and unbiased) estimate to $f_{\D}(A)$. Note that bags for different values of $\ell$ (on the same sampled positions) can be obtained by maintaining a table $T_{\ell}$ and a set $\mathcal{P}_{A,\ell}$ for each value $\ell$  of interest.

Note that \algname\ does not sample $m$ bags of \emph{exactly} $\ell$ positions each, since the number of occurrences of each position in $\D$ in the sample $P_\ell$ is sampled independently from a Poisson distribution, even if the expected number of total occurrences sampled from the algorithm is $m \ell$. However, the independent Poisson distributions used by \algname\ provide an accurate approximation of the random sampling of \emph{exactly}  $m \ell$ positions used in the analysis of Section~\ref{sec:sample_bags}.  In particular, this holds when one focuses on the events of interests for our approximation of Section~\ref{sec:sample_bags} (e.g., the event ``there exists a $k$-mer $A$ such that $|\E[\hat{f}_{P_\ell}(A)] - \hat{f}_{P_\ell}(A)| > \varepsilon / 2$"). In fact, a simple adaptation of a known result (Corollary 5.11 of~\cite{mitzenmacher2017probability}) on the relation between sampling with replacement and the use of independent Poisson distributions gives the following.
\begin{proposition}
Let $E$ be an event whose probability is either monotonically increasing or monotonically decreasing in the number of sampled positions. If $E$ has probability $p$ when the independent Poisson distributions are used, then $E$ has probability at most $2p$ when the sampling with replacement is used.
\end{proposition}
As a simple corollary, the output $\mathcal{O}$ features the guarantees of Proposition~\ref{thm:kmsamplesizevellpositions} with probability $\ge 1-  \delta'$, with $\delta'=2 \delta$.

\subsection{Improved Lower and Upper Bounds to $k$-mers Frequencies}
\label{sec:lower_upper_bounds}
Note that Proposition~\ref{thm:kmsamplesizevellpositions} guarantees that we can obtain upper and lower bounds to $f_\D(A)$ for every $A \in \fkm(\D,k,\theta)$ from the sample of bags of $\ell$ positions. These bounds are meaningful only in specific ranges of the frequencies; for example, the lower bound  from the third guarantee in Proposition~\ref{thm:kmsamplesizevellpositions} is meaningful when the frequency of $A$ is fairly low, i.e $f_\D(A) \approx 1 / \ell$, while for very frequent $k$-mers they could be a multiplicative factor $1/\ell$ away from than the correct value. For example, if a $k$-mer is very frequent and appears in all bags of $\ell$ $k$-mers in a sample $S$, its corresponding lower bound is still only $1 / \ell - \varepsilon/2$.

However, Proposition~\ref{thm:kmsamplesizevellpositions} can be  generalized to obtain tighter upper and lower bounds to the frequency of all $k$-mers. For given $\ell$, $\varepsilon$, and $\delta$, let $m$ as given in Proposition~\ref{thm:kmsamplesizevellpositions}. Note that the total number of $k$-mer's positions in the sample $P_\ell$ is $m\ell$.
Let $\mathcal{L}$ be a set of integer values $\mathcal{L} = \{ \ell_i \}$ with $\ell_i \in [1,m\ell], \forall i=0,\dots,|\mathcal{L}|-1$. Now, for every $\ell_i \in \mathcal{L}$, we can partition the \emph{same} $m\ell$ $k$-mers that are in $P_\ell$ into $m_i=m\ell/\ell_i$ partitions having size $\ell_i$. Let $P_{\ell_i}$ be such a random partition of such positions into $m_i$ bags of $\ell_i$ positions each. Note that each $P_{\ell_i}$ is a ``valid'' sample (i.e., a sample of independent bags of positions, each obtained by uniform sampling with replacement) for Proposition~\ref{thm:kmsamplesizevellpositions}, even if the $P_{\ell_i}$'s are not independent.  {\sloppy From each $P_{\ell_i}$, we define a maximum deviation $\varepsilon_i$ from Proposition~\ref{thm:kmsamplesizevellpositions} as $\varepsilon_i  = \frac{1}{\ell_i} \sqrt{\frac{2}{m_i} \left( \lfloor \log_2 (\min (2\ell_i , \sigma^k)) \rfloor + \ln \left(|\mathcal{L}| / \delta\right) \right)}$.}
We have the following result.

\begin{proposition}
\label{prop:bounds}
With probability at least $1-\delta$, for all $k$-mers $A$ simultaneously and for all the random partitions induced by $\mathcal{L}$ it holds 
\begin{itemize}
\item $f_\D(A) \ge \max\{ \hat{f}_{P_{\ell_i}}(A) - \varepsilon_i/2: i=0,\dots,|\mathcal{L}|-1 \}$;
\item $f_{\D}(A) \ge \max\{1- ( 1 - \ell (\hat{f}_{P_{\ell_i}}(A) - \varepsilon_i/2))^{1 / \ell}: i=0,\dots,|\mathcal{L}|-1$ and $\hat{f}_{P_\ell}(A) - \varepsilon_i/2 \ge 0 \} $;
\item $f_{\D}(A) \le\min\{1- ( 1 - \ell (\hat{f}_{P_{\ell_i}}(A) + \varepsilon_i/2))^{1 / \ell}: i=0,\dots,|\mathcal{L}|-1$ and $\hat{f}_{P_\ell}(A) + \varepsilon_i/2 \le 1/\ell \} $.
\end{itemize}
\end{proposition}

In our experiments, we use $\mathcal{L} = \{ \ell_i \}$ with $\ell_i = \ell / 2^i, \forall i \in [0 , \lfloor \log_{2}\ell \rfloor - 1]$; in this case, note that $P_{\ell_0}=P_{\ell}$. Using this scheme, we can compute upper and lower bounds for $k$-mers having frequencies of many different orders of magnitude, but any (application dependent) distribution can be specified by the user. These upper and lower bounds can be used to obtain different approximations of $\fkm(\D,k,\theta)$ with different guarantees. For example, by reporting all $k$-mers (and their frequencies) that have an upper bound $\ge \theta$, we have an approximation that guarantees that all $k$-mers $A$ with $f_{\D}(A) \ge \theta$ are in the approximation.

\section{Experimental Results}
\label{sec:experiments}
In this section we present the results of our experimental evaluation for \algname. 
Section \ref{sec:data} describes the datasets, our implementation for \algname\footnote{Available at https://github.com/VandinLab/SAKEIMA}, and the baseline for comparisons.
In Section \ref{sec:experimentsfrequents}, we report the results for computing the approximation of the frequent $k$-mers using \algname.
Section \ref{sec:experimentsdistances} reports the results of using our approximation to compute abundance-based and presence-based distances between metagenomic datasets. 

\subsection{Datasets and Implementation}
\label{sec:data}

We considered six datasets from the Human Microbiome Project (HMP)\footnote{https://hmpdacc.org/HMASM}, one of the largest publicly available collection of metagenomic datasets from high-throughput sequencing. In particular, we selected the three largest datasets of \texttt{stool} and the three largest of \texttt{tongue dorsum} (Table \ref{tab:data}). 
These datasets constitute the most challenging instances, due to their size, and provide a test case with different degrees of similarities among datasets.

\begin{table}
\caption{Datasets for our experimental evaluation. For each dataset $\D$ the table shows: the dataset name and site (\texttt{(s)} for \texttt{stool}, \texttt{(t)} for \texttt{tongue dorsum}); the total number $t_{\D,k}$ of $k$-mers ($k=31$) in $\D$; the number $|\D|$ of reads it contains; the maximum read length $\max_{n_i} = \max_i\{n_i | r_i \in \D\}$; the average read length avg$_{n_i} = \sum_{i=0}^{n-1} n_i/n$.}
\label{tab:data}
   \centering
\footnotesize{
   \begin{tabular}{|c|r|r|r|r|}
      \hline
      dataset & \multicolumn{1}{|c|}{$t_{\D,k}$}& \multicolumn{1}{|c|}{$|\D|$} & \multicolumn{1}{|c|}{$\max_{n_i}$} & avg$_{n_i}$ \\
       \hline
       \hline
      \texttt{SRS024388(s)}  & $7.92 \cdot 10^9$& $1.20 \cdot 10^8$ & 102 & 97.21 \\
       \hline
      \texttt{SRS011239(s)}  & $8.13 \cdot 10^9$& $1.24 \cdot 10^8$ & 102 & 96.69 \\
       \hline
      \texttt{SRS024075(s)}  & $8.82 \cdot 10^9$& $1.38 \cdot 10^8$ & 96 & 94.88 \\
       \hline
      \texttt{SRS075404(t)} & $7.75 \cdot 10^9$& $1.22 \cdot 10^8$ & 102 & 94.51 \\
       \hline
      \texttt{SRS062761(t)}  & $8.26 \cdot 10^9$& $1.18 \cdot 10^8$ & 101 & 101.00 \\
       \hline
      \texttt{SRS043663(t)}  & $9.15 \cdot 10^9$& $1.31 \cdot 10^8$ & 101 & 101.00 \\
      
      \hline
   \end{tabular}
  }
\end{table}

We implemented \algname\ in \texttt{C++} as a modification of Jellyfish~\cite{marccais2011fast} (the version we used is 2.2.10\footnote{https://github.com/gmarcais/Jellyfish}), a very popular and efficient algorithm for exact $k$-mer counting. Doing so, our algorithm enjoys the succinct counting data structure provided by Jellyfish publicly available implementation. We remark that our sampling-based approach can be used in combination with any other highly tuned method available for exact, approximate, and parallel $k$-mer counting. For this reason, we only compare \algname\ with the exact counting performed by Jellyfish, since they share the underlying characteristics, allowing us to evaluate the impact of \algname\ sampling strategy.
We did not include the time to compute $t_{\D,k}$ in our experiments since it was always negligible (i.e., less than 2 minutes) w.r.t. the time for counting $k$-mers.

For the computation of the abundance-based distances
from the $k$-mer counts of two dataset, we implemented in \texttt{C++} a simple algorithm  that loads the counts of one dataset in main memory and then performs one pass on the counts of the other dataset, producing the distances in output.
We executed all our experiments on the same machine with $512$ GB of RAM and a 2.30 GHz Intel Xeon CPU, compiling both implementations with \texttt{g++ 4.9.4}. \algname\ can be used in combination with more efficient algorithms and implementations for the computation of these (and other) distances~\cite{benoit2016multiple}, resulting in speed-ups analogous to the ones we present below.
For all the experiments of \algname, given $\theta$ and a dataset $\D$, we fixed the parameters $\delta = 0.1$, $\varepsilon = \theta - 2/t_{\D,k}$, $m = 100$, and we fix $\ell$ to the minimum value satisfying the $\varepsilon$-approximation. For all the experiments we have $\ell$ close from below to $1/\theta$. For all the metrics we considered, we report the results for one random run.

\subsection{ Approximation of the Frequent $k$-mers}

\label{sec:experimentsfrequents}
We fixed $k=31$, and we compared \algname\ with the exact counting of all $k$-mers (from Jellyfish) in terms of:
\begin{inparaenum}[(i)]
\item running time\footnote{Every instance of \algname\ and Jellyfish was executed with $1$ worker, i.e., sequentially. 
Note that the Poisson approximation employed by \algname\ allows multiple workers to independently process the input $k$-mers, therefore \algname\ can be used in a parallel scenario. We will investigate the impact of parallelism in the extended version of this work.
}, including, for both algorithms, the time required to write the output on disk;
\item memory requirement.
\end{inparaenum}
We also assessed the accuracy of the output of \algname.

\begin{figure}[h]
  \centering
  \includegraphics[width=0.9\textwidth]{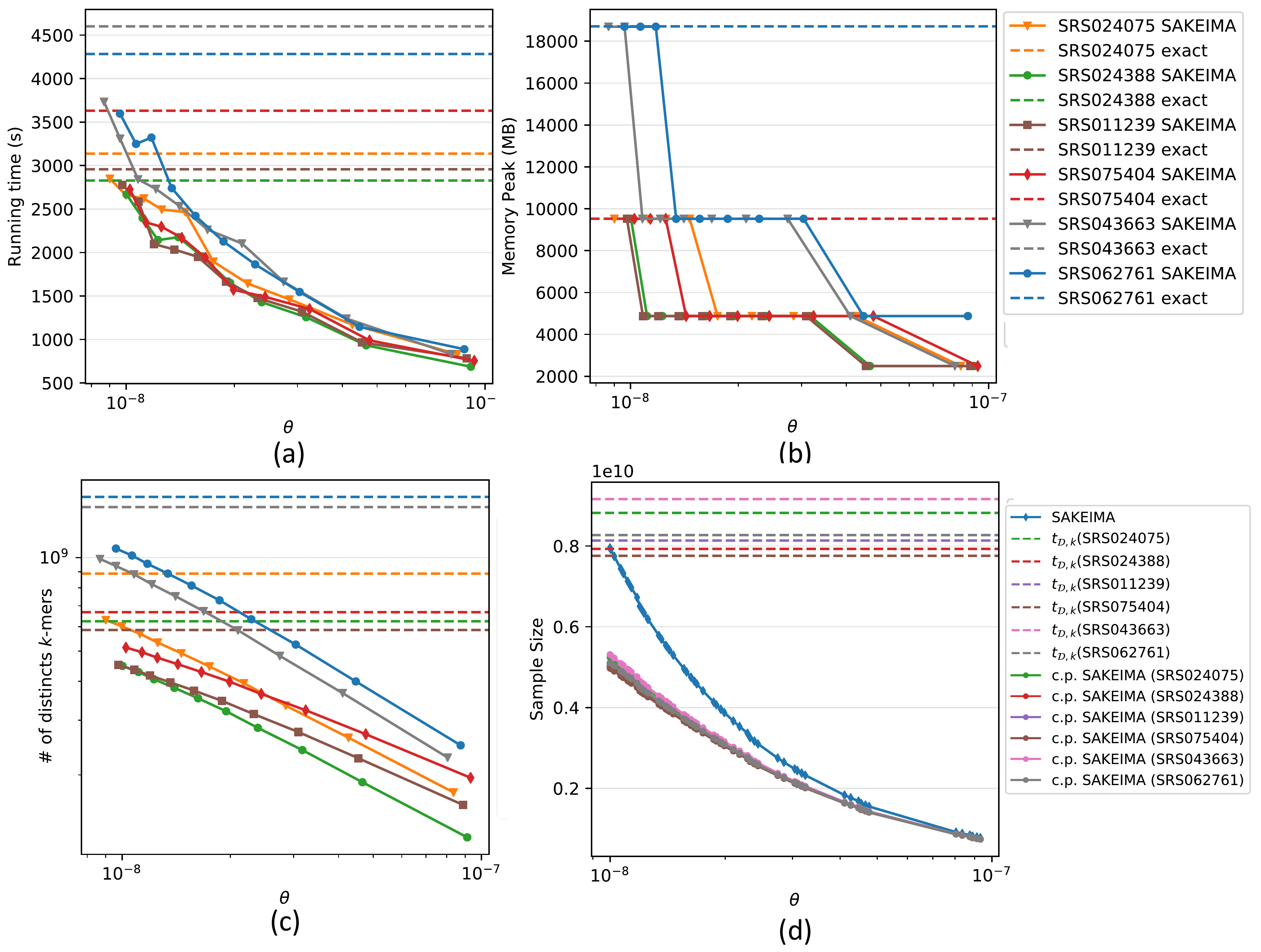}
  \caption{Running time, memory requirements, and number of distinct $k$-mers counted, for \algname\ and exact counting as function of $\theta$. (a) Running time. (b) Memory requirement. (c) Number of distinct $k$-mers counted. (d) Sample sizes of \algname, total size $t_{\D,k}$ of the datasets, and number (c.p.) of dataset's distinct covered positions  (i.e., included in \algname's sample), as function of $\theta$.}
   \label{fig:fig1_combined}
\end{figure}

Figure~\ref{fig:fig1_combined} shows the running times and the peak memory as function of $\theta$. Note that for the exact counting algorithm these metrics do not depend on $\theta$, since it always counts all $k$-mers. \algname\ is always faster than the exact counting, with a difference that increases when $\theta$ increases and a speed-up around $2$ even for $\theta = 2 \cdot 10^{-8}$. The memory requirement of \algname\ reduces when $\theta$ increases, and for $\theta = 2 \cdot 10^{-8}$ it is half of the memory required by the exact counting. This is due to \algname's sample size
being much smaller than the dataset size (Figure~\ref{fig:fig1_combined}(d)), therefore a large portion of extremely low frequency $k$-mers are naturally left out from the random sample and do not need to be accounted for in the counting data structure, as confirmed by counting the number of \emph{distinct} $k$-mers that are inserted in the counting data structure by the two algorithms (Figure~\ref{fig:fig1_combined}(c)). (The difference between the memory requirement and the number of distinct $k$-mers is given by Jellyfish's strategy to doubles the size of the counting data structure when it is full.)

\begin{figure}[h]
  \centering
  \includegraphics[width=0.9\textwidth]{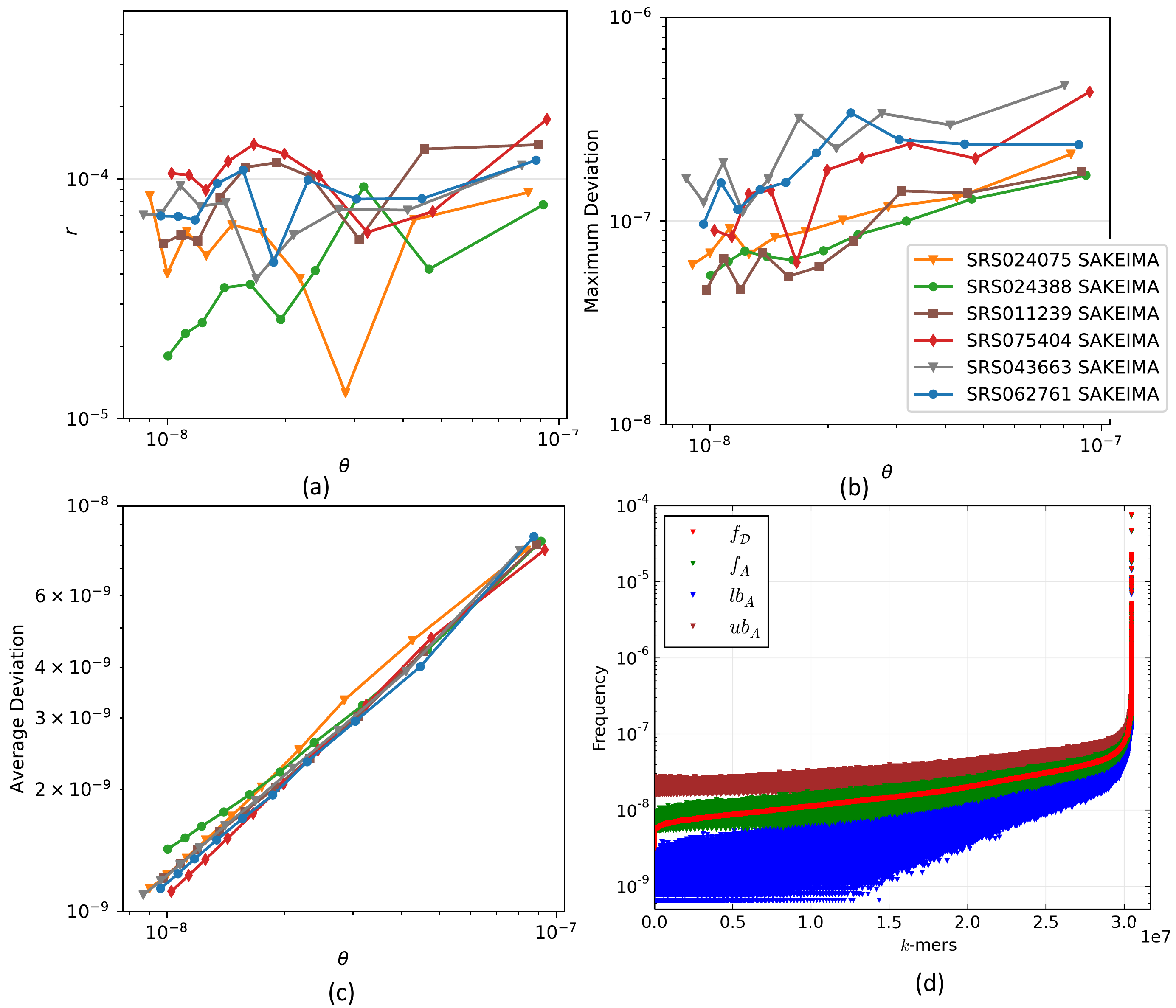}
  \caption{Quality of the approximation of $\fkm(\D,k,\theta)$ produced by \algname. (a) False negative rate, i.e., the fraction $r$ of $k$-mers in $\fkm(\D,k,\theta)$ not reported by \algname. (b) Maximum deviation $|f_A - f_{\D}(A)|$ of the estimates reported by \algname\ for various $\theta$. (c) Average value of $|f_A - f_{\D}(A)|$ for the $k$-mers $A$ reported by \algname\ for various $\theta$. (d) Frequencies and bounds for dataset \texttt{SRS062761} and $\theta = 10^{-8}$ shown for $k$-mers sorted in increasing order of exact frequencies. Red: exact frequencies $f_\D(A)$. Green: estimate $f_A$ of $f_\D(A)$ from \algname. Blue: lower bound $lb_A$ to $f_\D(A)$ from \algname. Brown: upper bound $ub_A$ to $f_\D(A)$ from \algname.}
   \label{fig:fig2_combined}
\end{figure}

In terms of quality of the approximation, the output of \algname\ satisfied the guarantees given by Proposition~\ref{thm:kmsamplesizevellpositions} for all runs of our experiments, therefore with probability higher than $1 - \delta$. While \algname\ may incur in false negatives, its false negative ratio (i.e., the fraction of $k$-mers in $\fkm(\D,k,\theta)$ not reported by \algname) is always $\le 3 \cdot 10^{-4}$ (Figure~\ref{fig:fig2_combined}(a)),
 even if the sampling technique of Section~\ref{sec:sample_bags} does not provide rigorous guarantees on such
quantity. Therefore \algname\ is very effective in reporting almost all frequent $k$-mers. As mentioned in Section~\ref{sec:lower_upper_bounds}, \algname\ can be easily modified so to report all frequent $k$-mers in output, even if at the cost of reporting also more $k$-mers with frequency between $\theta-\varepsilon$ and $\theta$. In addition, the estimated frequencies $f_A$ reported by \algname\ are always close to the true values $f_{\D}(A)$, with a small maximum deviation $|f_A - f_{\D}(A)|$ (Figure~\ref{fig:fig2_combined}(b)), and an even smaller average deviation (Figure~\ref{fig:fig2_combined}(c)). In addition, the upper and lower bounds computed as in Section~\ref{sec:lower_upper_bounds} provide small confidence intervals always containing the value $f_{\D}(A)$ (e.g., Figure~\ref{fig:fig2_combined}(d) for dataset \texttt{SRS062761}), and could be used to obtain sets of $k$-mers with various guarantees from the sample used by \algname.

\subsection{Application to Metagenomics: Computation of Ecological Distances}
\label{sec:experimentsdistances}
We evaluate the use of \algname\ to speed up the computation of commonly used $k$-mer based ecological distances~\cite{benoit2016multiple} between datasets of Next-Generation Sequencing (NGS) reads. We present results for the Bray-Curtis distance; analogous results hold for other distances and will be presented in the full version of this extended abstract.

We first investigated how the distances change when those are computed considering only the \emph{frequent} $k$-mers (w.r.t. a frequency threshold $\theta$) instead that the full spectrum of $k$-mers appearing in the data. Therefore, given a pair of datasets $\D_1$ and $\D_2$ and $\theta$, we computed the sets $\mathcal{O}_1 = \fkm(\D_1,k,\theta)$ and $\mathcal{O}_2 = \fkm(\D_2,k,\theta)$ using Jellyfish and then computed a generalized version of the distances for all pairs of datasets we used for our experiments. For the Bray-Curtis distance, this generalization is defined as: $BC(\D_1 , \D_2 , \mathcal{O}_1 , \mathcal{O}_2) = 1 - 2 \frac{\sum_{A \in \mathcal{O}_1 \cap \mathcal{O}_2} \min\{ o_{\D_1}(A) , o_{\D_2}(A) \} }{\sum_{A \in \mathcal{O}_1 } o_{\D_1}(A) + \sum_{A \in  \mathcal{O}_2} o_{\D_2}(A)}$.

\begin{figure}
  \centering
  \includegraphics[width=0.9\textwidth]{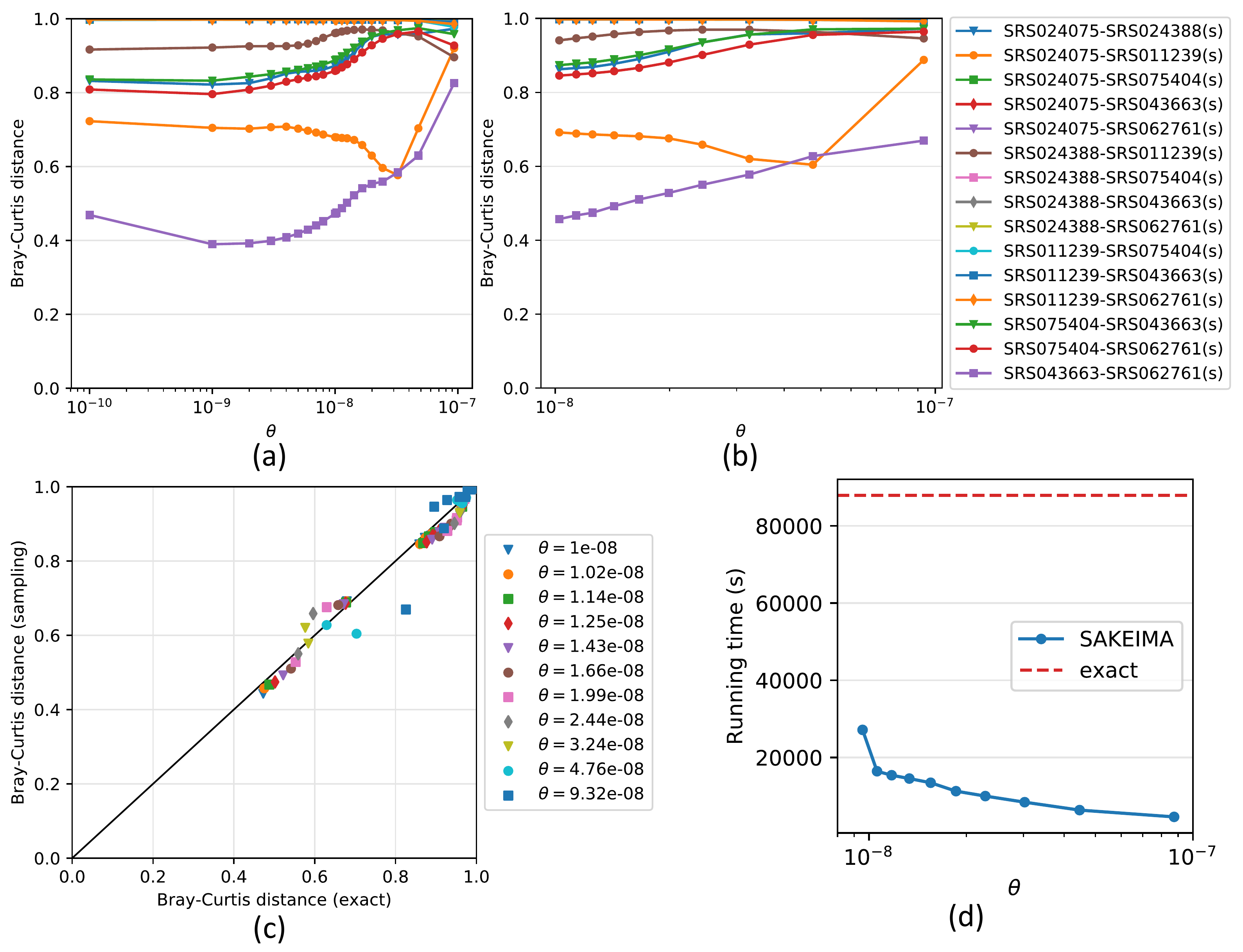}
  \caption{Results for Bray-Curtis (BC) distances of metagenomic datasets. (a) BC distance computed using $k$-mers with frequency $\ge \theta$.  (b) BC distances computed using the approximation of $k$-mers with frequency $\ge \theta$ from \algname. (c) Comparison of the BC distance using all $k$-mers with exact counts and the approximation of frequent $k$-mers by \algname. (d) Total time required by \algname\ and the exact approach to find frequent $k$-mers and compute all distances between datasets as a function of $\theta$.}
   \label{fig:fig3_combined}
\end{figure}

Note that when $\theta \leq 10^{-10}$ then $\fkm(\D,k,\theta)$ coincides with the set of \emph{all} $k$-mers, for any of the datasets we tested. The results (Figure~\ref{fig:fig3_combined}(a)) show that for $\theta$ up to $5\times10^{-8}$ the values of the distances are fairly stable and therefore one can use only frequent $k$-mers for such values of $\theta$ to compute the distances, and that for $\theta$ up to $10^{-7}$ the relation between distances of different pairs of datasets are almost always conserved. We underline that the exact counting approach needs to count \emph{all} the $k$-mers and only afterwards can filter the infrequent ones before writing them to disk to compute $\fkm(\D,k,\theta)$.
We then used \algname\ to extract approximations (of $k$-mers and their frequencies) of $\fkm(\D_1,k,\theta)$ and $\fkm(\D_2,k,\theta)$ and used such approximations to compute the distances among datasets (Figure~\ref{fig:fig3_combined}(b)). Strikingly, the distances computed from the output of \algname\ are very close to their exact variant (Figure~\ref{fig:fig3_combined}(c)). Interestingly this holds also for the Jaccard distance,
a presence-based distance that does not depend neither on $k$-mer abundances nor on $k$-mer ranking by frequencies (detailed results will be provided in the full version of this extended abstract).

We then compared, for different values of $\theta$, the total running time required to compute the approximations of the frequent $k$-mers using \algname\ for all datasets in Table~\ref{tab:data} and all distances among such datasets using \algname\ approximations with the running time required when the exact counting algorithm is used for the same tasks. \algname\ reduces the computing time by more than $75\%$ (Figure~\ref{fig:fig3_combined}(d)). This result comes from both the efficiency of \algname\ and from the fact that by focusing on the the most frequent $k$-mers we greatly reduce the number of distinct $k$-mers that need to be processed for computing the distances. Therefore \algname\ can be used for a very fast comparison of metagenomic datasets while preserving the ability of distinguishing similar datasets from different ones.

\section{Conclusion}

We presented \algname, a sampling-based algorithm to approximate frequent $k$-mers and their frequencies with rigorous guarantees on the quality of the approximation. We show that \algname\ can be used to speed up the analysis of large high-throughput sequencing metagenomic datasets, in particular to compute abundance-based distances among such datasets. Interestingly \algname\ allows to compute accurate approximations also for presence-based distances (e.g., the Jaccard distance), even if for such distances other, potentially faster, tools~\cite{ondov2016mash} are available. \algname\ can be combined with any highly optimized method that counts all $k$-mers in a set of strings, including recent parallel methods designed for comparative metagenomics~\cite{benoit2016multiple}. While we presented results for $k$-mers from datasets of short reads, \algname\ can also be used for the analysis of spaced seeds~\cite{bvrinda2015spaced}, large datasets of long reads, and whole genome sequences.

\bibliographystyle{plainurl}

\end{document}